\documentclass[11pt]{article}
\usepackage[margin=1in]{geometry}
\usepackage{natbib}

\usepackage[textwidth=1.4cm, textsize=small, colorinlistoftodos]{todonotes}

\usepackage{amsmath}
\usepackage{amssymb}
\usepackage{graphicx}
\usepackage[utf8]{inputenc}

\usepackage{enumerate}
\usepackage{verbatim}

\usepackage{hyperref}
\usepackage[capitalize, noabbrev]{cleveref}

\newcommand{\rst}{\emph{Roman }}

\begin{document}

\noindent
\textbf{\large Roman CCS White Paper}

\vspace{0.5cm}

\begin{center}
{\bf
    Tracing stellar mass assembly and emerging quiescence at cosmic noon: the case for deep imaging with \emph{all} of Roman's wide filters in the HLTDS}
\end{center}

\vspace{0.8cm}

\noindent {\bf Core Community Survey:} High Latitude Time Domain Survey (HLTDS)\\

\noindent {\bf Scientific categories:} galaxies, stellar physics and stellar types\\

\noindent  {\bf Additional scientific keywords:} deep fields, supernova cosmology, dark energy\\

\noindent {\bf Submitting author:}\\
Bhavin A.~Joshi, Johns Hopkins University (bjoshi5@jhu.edu)\\

\noindent {\bf Contributing authors:}\\
Louis-Gregory Strolger, Space Telescope Science Institute (strolger@stsci.edu)\\
Sebastian Gomez, Space Telescope Science Institute (sgomez@stsci.edu)\\
Benjamin Rose, Baylor University (Ben\_Rose@baylor.edu)\\


\vspace{1em}
\noindent
\textbf{Abstract:}\\
We present arguments for including observations with all of the Wide Field Instrument imaging filters, with the exception of F146, within the Nancy Grace Roman Space Telescope (\emph{Roman}) High Latitude Time Domain Survey (HLTDS). Our case is largely driven by the extragalactic deep field science that can be accomplished with HLTDS observations and also by the improvements in type Ia supernova (SN~Ia) cosmology systematics that a wide wavelength coverage affords.

\clearpage

\section{Motivation}
It is well established that most galaxies in the Universe experienced a peak in star-formation around $z{\sim}2\text{--}3$ \citep{Madau2014}; a period known as cosmic noon. It is also well established, through large photometric samples and spectroscopic confirmations, that a significant fraction of massive galaxies at these redshifts were already quiescent by this time \citep{Straatman2014, Glazebrook2017, Belli2019, Valentino2020}. Galaxy redshifts and stellar masses are two key measurements that are critical to this understanding. They can be robustly determined from photometry; however, photometry that is not deep enough or does not cover 4000\AA/Balmer breaks or the restframe near-IR, can significantly bias either measurement \citep[e.g.,][]{Pforr2012, Mobasher2015}.
Some essential questions still remain in galaxy evolution -- what processes are responsible for quenching and maintaining quiescence? \citep[e.g.,][]{Whitaker2013, Feldmann2015, Man2018} Since the timescales are vastly different it is unlikely that a single process is responsible. What processes drive the scatter in the star-forming main sequence of galaxies? \citep{Whitaker2012, Speagle2014, Schreiber2015, Donnari2019} Additionally, while it is suspected that Lyman continuum from low mass galaxies dominated the ionizing budget during and prior to the epoch of reionization \citep[$z{\gtrsim}6$;][]{Bouwens2012, Stark2016} and that more massive galaxies and AGN were responsible for maintaining an ionized IGM later on \citep[e.g.,][]{Steidel2018, Smith2020, Jones2021}, the details of this process remain unclear.
We show that \rst High Latitude Time Domain Survey (HLTDS) observations can be easily adapted to address these questions.

The reference survey for the HLTDS presented in \citet{Rose2021} is divided into a wide and a deep tier with a 75:25 split between imaging and prism spectroscopy (\cref{tab:ref}). The imaging is spread out such that the wide tier is covered by F062, F087, F106, and F129, whereas the deep tier is covered by F106, F129, F158, and F184. \emph{However, neither tier receives coverage by more than four filters, and the reddest filter F213 is not utilized at all.} We present options (\cref{tab:choices}) that update the reference survey (in either the deep or the wide tier or both) to address fundamental questions about galaxy evolution, significantly increasing the legacy value of the HTLDS beyond supernova cosmology. This is important because in the absence of a planned \rst deep field the HLTDS deep tier will serve as the de facto deep field for extragalactic science with \emph{Roman}. In addition, this broad wavelength coverage lays the foundation for future general astrophysics programs with \rst to benefit from an ultra-deep tier embedded within the HLTDS deep tier that consists of one or several pointings.

\begin{table}[t!]
    \centering
    \begin{tabular}{c c p{4cm} c c c}
    \hline
    Tier & Filters & Exposure time + overhead [s] & Pointings & Area [deg$^2$] & Time/Visit [hr] \\
    \hline
  Wide  & F062;F087;F106;F129 & 160;100;100;100 + 70$\times$4 & 68 & 19.04 & 14.0 \\
  Deep  & F106;F129;F158;F184 & 300;300;300;900 + 70$\times$4 & 15 & 4.20 & 8.5\\
  \hline
    \end{tabular}
    \caption{The imaging part of the reference survey reproduced from \citet{Rose2021}. The remainder of the 30 hr per visit time is devoted to prism spectroscopy.}
    \label{tab:ref}
\end{table}

\cref{tab:ref} shows the design parameters for the reference survey (reproduced from \citealt{Rose2021}). The updates presented here are computed while maintaining the assumption of a total survey duration of 2 years and a 5 day cadence.
Case (i) shown expands the deep tier of the reference survey by including F062, F087, and F213 coverage and thereby providing {\bf data in 7 filters from $\sim$0.6--2.1$\mu m$ over 4 deg$^2$ complete to a depth of $\sim$29 AB mag for point sources}. This represents a increase of $\sim$7.1 hr in observing time per visit (over the 30 hr total exposure time for the reference survey; including overheads). 
Case (ii) shown expands the wide tier of the reference survey by including F158, F184, and F213 coverage and again providing {\bf data in 7 filters from $\sim$0.6--2.1$\mu m$ over 19 deg$^2$ complete to a depth of $\sim$27--28 AB mag (filter dependent) for point sources}. This similarly represents a increase of $\sim$9.6 hr in observing time per visit.
While the F213 filter is not actively cooled and therefore has a higher thermal background we have still included this filter in our survey options given the importance of measuring restframe near-IR flux for stellar mass (and also redshift) measurements.

\begin{table}
    \centering
    \begin{tabular}{c p{6cm} c c c}
    \hline
Tier & Filter (Exposure time [s]) & Time/visit [hr] & Depth [AB mag] & Area [deg$^2$]\\
    \hline
  Deep  & F062 (300), F087 (300), F106 (300), F129 (300), F158 (300), F184 (900), F213 (900) & 37.13 & $\sim$29 & $\sim$4\\
  Wide  & F062 (160), F087 (100), F106 (100), F129 (100), F158 (100), F184 (100), F213 (100) & 39.63 & $\sim$28 & $\sim$19\\
  \hline
    \end{tabular}
    \caption{Choices for extending the HLTDS reference survey }
    \label{tab:choices}
\end{table}

\section{Science Goals}
\label{sec:goals}

\begin{figure}
    \centering
    \includegraphics[width=0.75\textwidth]{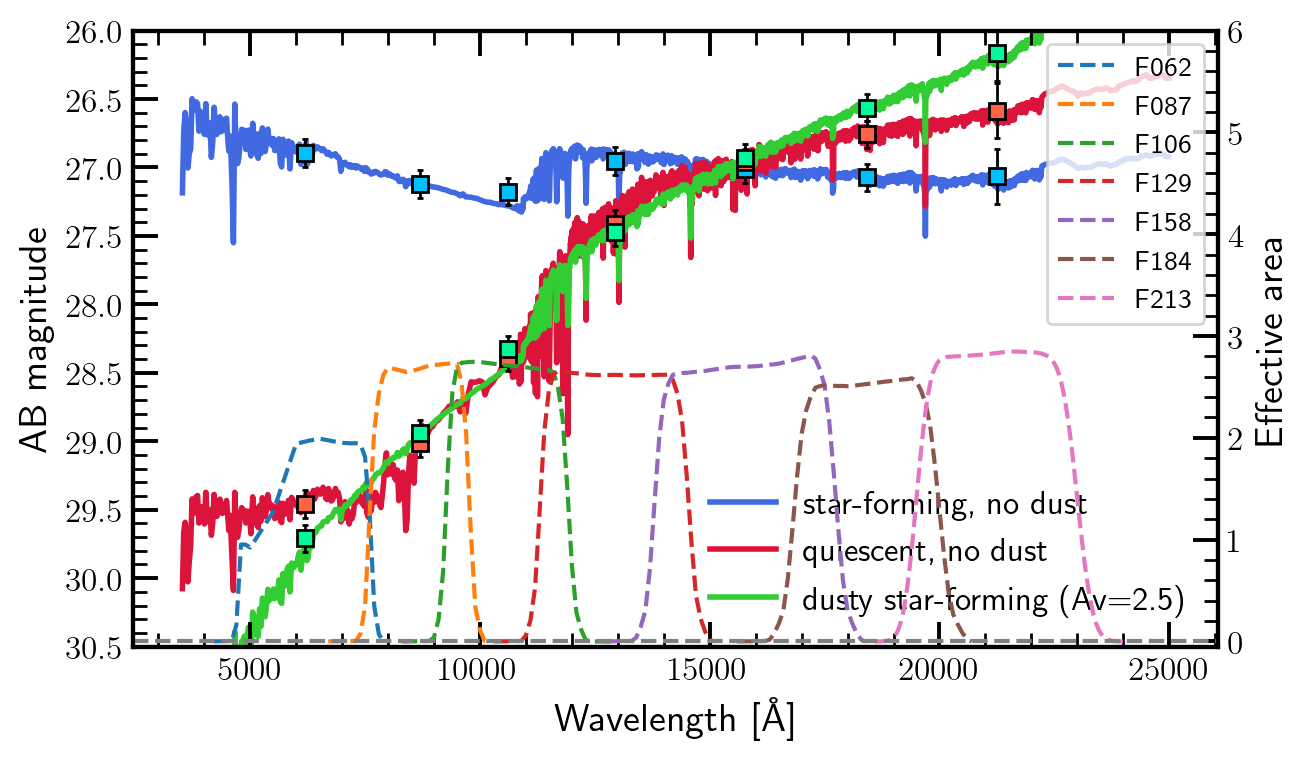}
    \caption{Representative spectral energy distributions (SEDs) at cosmic noon (assumed $z{=}2$ here) and their photometry through Roman WFI bandpasses (excluding F146). All SEDs are normalized to an AB magnitude of 27 in the F158 bandpass. Roman filter curve effective areas are also shown at the bottom of the figure with the values on the right ordinate. We assume a 5$\sigma$ measurement significance in the F213 filter and 10$\sigma$ for all other filters. The SEDs are simulated using the FSPS package \citep{Conroy2009, Conroy2010}.}
    \label{fig:seds}
\end{figure}

\noindent {\bf Star-formation and quiescence at cosmic noon:} Much of the advance in our knowledge of star-formation and quiescence at $z{>}1$ has come through deep multi-wavelength surveys with HST -- CANDELS \citep{Grogin2011, Koekemoer2011}, GOODS \citep{Giavalisco2004}, Frontier Fields \citep{Lotz2017}, CLASH \citep{Postman2012}. The galaxy stellar mass function (SMF) which is the number density of galaxies as a function of stellar mass traces the growth of baryonic matter over time. The star-forming main sequence of galaxies, which is the relationship between the star formation rates (SFR) and the stellar masses of star-forming galaxies, and the SMF provide means to study the assembly of stellar mass, the efficiency of star-formation over redshift (and over stellar mass), and the emergence of quiescence. To enable measurements of stellar masses at the the high mass end of the SMF for the most massive, rarest galaxies and at the fainter, low mass end (sub-M*) requires \rst sensitivity along with its wide FOV to overcome cosmic variance (which has been the disadvantage for HST extragalactic deep fields so far).\\
\cref{fig:seds} shows representative SEDs of three example galaxies at cosmic noon and their inferred photometry through \rst WFI bandpasses (excluding the extremely wide F146 filter). In an HLTDS design where the deep and wide tiers are each covered by four bandpasses, this figure highlights the difficulty that we would face in trying to distinguish between the SEDs. Note the striking similarly between the colors of the quiescent SED and the dusty star-forming SED, in the central four filters, underscoring the age-dust degeneracy. Only the photometry at the bluest and the reddest end is helpful in trying the tell these SEDs apart. This figure therefore also shows the inadequacy of a four filter SED to infer galaxy properties (including SN~Ia host galaxies). Coverage from all seven (or at least six) WFI filters therefore is necessary to distinguish between stellar populations with similar colors and correctly infer their properties.\\

\begin{figure}[ht!] 
	\includegraphics[width=0.5\textwidth]{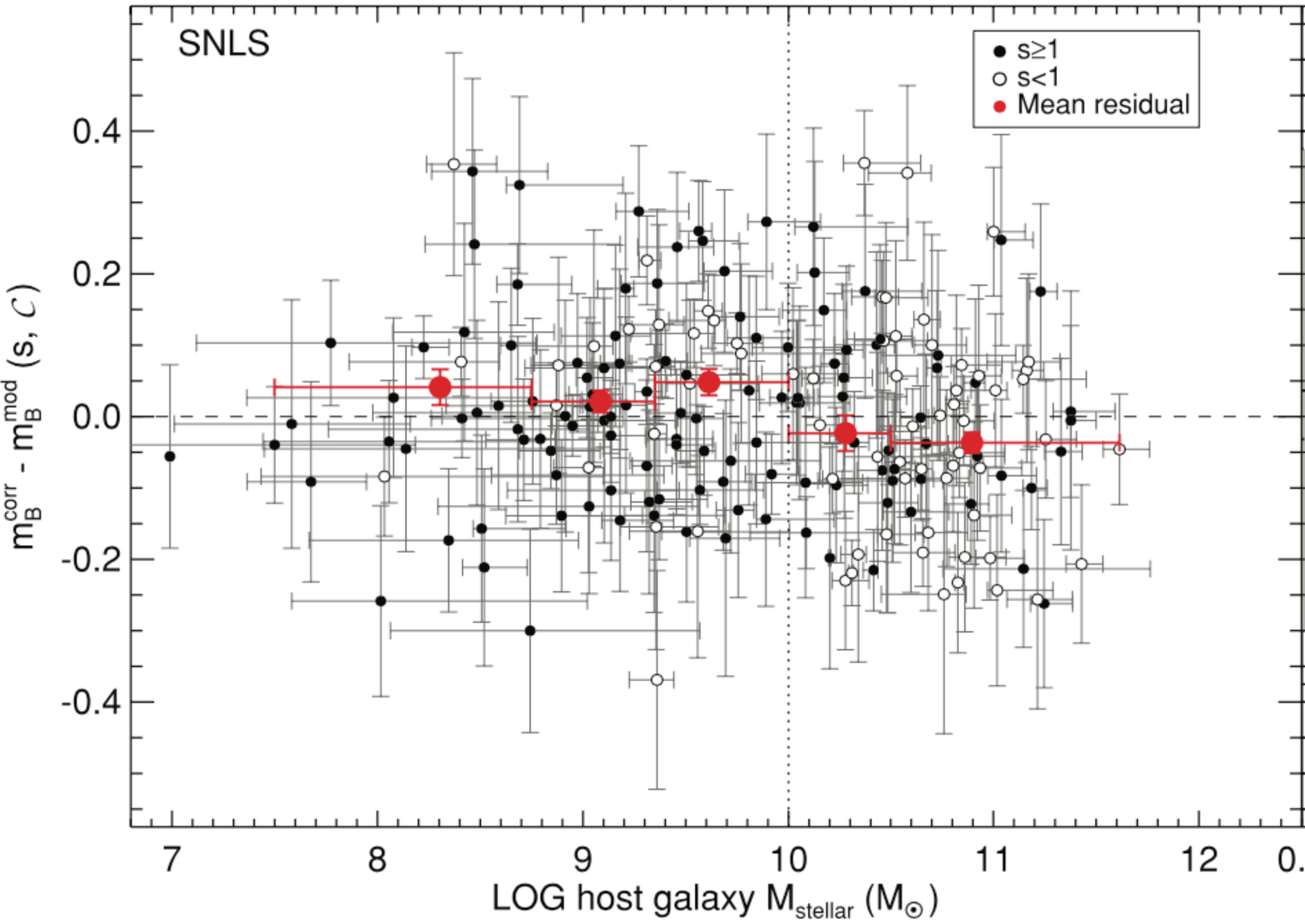}
	\includegraphics[width=0.5\textwidth]{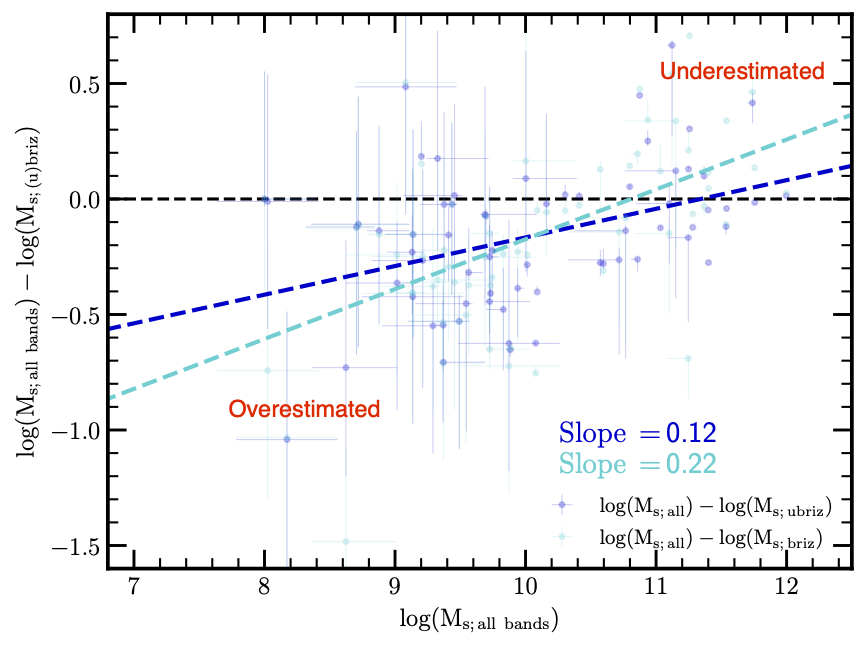}
   \caption{\footnotesize (\textit{Left}) Hubble residuals as a function of host galaxy stellar mass, reproduced from \citet{Sullivan2010}, for SNe host galaxies from the Supernova Legacy Survey. The red points are mean residuals in bins of stellar mass. (\textit{Right}) The distribution of differences of host galaxy stellar mass estimates for our test sample from CANDELS GOODS-N and GOODS-S. The stellar mass estimates from using the full filter set ($u$ to 3.6$\mu m$) vs a restricted photometric set of $(u)griz$ are compared. The dark/light blue points show the differences when the $u$ band is included/excluded. The best fit lines to the distributions are also shown in the same color as the points. Regions where masses are underestimated or overestimated are labeled.}
   \label{fig:resid_mass}
\end{figure}

\noindent {\bf Improved characterization of the mass-step systematic:} The ``mass-step'' systematic effect observed between Hubble residuals of SNe~Ia and the stellar masses of their host galaxies \citep{Sullivan2010, Kelly2010, Lampeitl2010} is routinely corrected for in cosmology analyses. However, typical SN cosmology analyses use a patchwork of optical photometry (and sometimes near-IR; from multiple surveys/instruments) to estimate galaxy stellar masses. This is potentially problematic given that lack of restframe near-IR flux measurements can lead to significant biases in stellar masses.\\
With a test sample of 66 SN~Ia host galaxies from CANDELS we show how stellar mass estimates are affected by lack of photometry in the restframe UV and especially in the restframe near-IR (\cref{fig:resid_mass}). This sample of SN~Ia hosts from CANDELS is chosen because it is representative of the redshift range at which \rst will observe most of its SNe~Ia ($0.5 \lesssim z \lesssim 3$). We compare the stellar masses inferred from SED fitting (through \texttt{Prospector}; \citealt{Leja2017, Johnson2021}) when using the ``full UV to near-IR'' filter set ($u$ to 3.6$\mu m$) vs a restricted photometric set of $(u)griz$. The biases introduced by employing a restricted set of photometry are typically 0.5 dex (overestimates) and 0.2 dex (underestimates) (\cref{fig:resid_mass} right panel).\\

\noindent {\bf Improved photometric redshifts:} \cref{fig:seds} also illustrates the difficulty of attempting to measure a photometric redshift with 4 bandpasses. It is also known from previous studies that $\sigma_z{\sim}0.05(1+z)$ for photometric redshifts determined from fewer optical only bands. However, the redshift uncertainty for the cosmology sample of supernovae from \rst is required to be at most at the $\sigma_z{\sim}0.001\text{--}0.002(1+z)$ level (at $z{\leq}1.5$; \rst science requirement SN 2.0.4). Given that most SN~Ia observed by \rst will not have a spectroscopic redshift this increases the demands on photometric calibration and on the ability to infer either a SN or host redshift from the photometry. 

\section{Trade offs}
The following HTLDS parameters are important for the science goals presented here: field location and overlap with existing observations, total coadded depth, and broad/restframe near-IR wavelength coverage. However, for example the cadence that is critical for SN~Ia light curve sampling is not so important for this science. The extremely wide F146 bandpass is also unlikely to be helpful for this science.

\bibliographystyle{aasjournal}
\bibliography{references}

\end{document}